\documentclass[%
reprint,
superscriptaddress,.
amsmath,
amssymb,
aps,
longbibliography,
floatfix,
]{revtex4-2}

\usepackage{mathrsfs,braket}
\usepackage{amssymb, amsbsy, amsmath, latexsym, dsfont, array, layout, graphicx,mathrsfs,color, bm,mathtools}
\usepackage{physics}
\usepackage[dvipsnames]{xcolor}
\usepackage[unicode=true,breaklinks=true,colorlinks=true]{hyperref}
\usepackage{bigints}
\usepackage[T1]{fontenc}
\usepackage{newtxtext}
\usepackage[utf8]{inputenc}
\usepackage{upgreek}
\usepackage{siunitx}

\hypersetup{citecolor=blue,urlcolor=blue,linkcolor=blue,filecolor=blue}
\newcommand{\eq}[1]{Eq.~\eqref{#1}}
\newcommand{\fig}[1]{Fig.~\ref{#1}}

\newcommand{\figsub}[2]{Fig.~\hyperref[#1]{\ref*{#1}(#2)}}
\newcommand{\fsub}[2]{\hyperref[#1]{\ref*{#1}(#2)}}
\newcommand{\figgsub}[2]{Figure~\hyperref[#1]{\ref*{#1}(#2)}}

\usepackage[normalem]{ulem}

\begin{document}

\title{Transient dynamics of parametric driving for single-electron image current detection in a Paul trap}

\author{Baiyi Yu}
\email{baiyi_yu@berkeley.edu}
\affiliation{Department of Physics, University of California, Berkeley, CA 94720, USA}
\affiliation{Challenge Institute for Quantum Computation, University of California, Berkeley, CA 94720, USA}
\affiliation{School of Physics, Huazhong University of Science and Technology, Wuhan 430074, China}
\author{Andris Huang}
\author{Isabel Sacksteder}
\author{Hartmut Haeffner}
\email{hhaeffner@berkeley.edu}
\affiliation{Department of Physics, University of California, Berkeley, CA 94720, USA}
\affiliation{Challenge Institute for Quantum Computation, University of California, Berkeley, CA 94720, USA}

\begin{abstract}
Nondestructive detection of single-electron motion is crucial for quantum information processing with electrons trapped in Paul traps.
The standard approach in Penning traps is to detect the image current induced on the trap electrodes by the electron's oscillatory motion.
However, applying this approach in Paul traps for single electrons is currently hindered by motional frequency fluctuations arising from trap anharmonicities and instabilities in the rf trapping field.
In this work, we propose a robust detection scheme exploiting the transient dynamics of parametric driving to overcome these limitations. 
Distinct from traditional steady-state approaches, our method focuses on the transient regime to break the temporal constraints imposed by steady-state assumptions, thereby enabling fast readout.
We show that a controlled ramp of the parametric drive effectively locks the frequency of the electron motion in the transient regime, rendering the signal highly resilient to realistic experimental noise and inherent micromotion.
This work paves the way for the experimental realization of nondestructive detection of single-electron motion in Paul traps.
 
\end{abstract}

\date{\today}

\maketitle
\section{Introduction}
Electrons trapped in vacuum have a long history of precision measurement, most notably within Penning traps~\cite{brownGeoniumTheoryPhysics1986, vandyckNewHighprecisionComparison1987, mittlemanBoundMathitCPTLorentz1999, peilObservingQuantumLimit1999, haffnerDoublePenningTrap2003, hannekeNewMeasurementElectron2008, fanMeasurementElectronMagnetic2023, fanQuantumlogicSpectroscopyElectron2025}.
Owing to their simple two-level internal spin states and the potential for exceptionally low noise from the trapping environments, they have recently emerged as strong candidates for quantum information processing (QIP) through architectures such as electrons trapped on liquid helium~\cite{platzmanQuantumComputingElectrons1999, wallraffStrongCouplingSingle2004, schusterProposalManipulatingDetecting2010, yangCouplingEnsembleElectrons2016, kawakamiImageChargeDetectionRydberg2019, kawakamiBlueprintQuantumComputing2023a, jenningsQuantumComputingUsing2024, beysengulovCoulombInteractionDrivenEntanglement2024, castoriaSensingControlSingle2025}, solid neon~\cite{zhouSingleElectronsSolid2022, zhouElectronChargeQubit2023, kanaiSingleElectronQubitsBased2024,  liElectronChargeCoherence2025,liNoiseresilientSolidHost2025,liCoherentManipulationInteracting2025}, and in Paul traps~\cite{daniilidisQuantumInformationProcessing2013,  matthiesenTrappingElectronsRoomTemperature2021,carneyTrappedElectronsIons2021, yuFeasibilityStudyQuantum2022, yuEngineeringArtificialAtomic2024, yuStrongCoherentIonelectron2024a, huangNumericalInvestigationsElectron2025, taniguchiImageCurrentDetection2025a, mikhailovskiiTrappingElectrons$^40mathrmCa^+$2026, laustiRoadmapPlanarElectronion2025}.
Among these approaches, the pure vacuum environment of the Paul trap is particularly attractive as it can directly adopt the mature control techniques~\cite{ospelkausMicrowaveQuantumLogic2011, srinivasHighfidelityLaserfreeUniversal2021} and the scalable architectures developed for trapped ions~\cite{haffnerQuantumComputingTrapped2008, pinoDemonstrationTrappedionQuantum2021, mosesRaceTrackTrappedIonQuantum2023, malinowskiHowWire$1000$Qubit2023}. 
Moreover, the charge-to-mass ratio of the electron is typically four orders of magnitude larger than that of commonly trapped ions, which speeds up many QIP-related operations by two orders of magnitude compared to their ion-based counterparts~\cite{yuFeasibilityStudyQuantum2022}.

One of the key components for QIP using electrons in Paul traps is the nondestructive detection of the state of a single electron using the image current~\cite{pengSpinReadoutTrapped2017}.
When an electron oscillates, it induces image currents in nearby electrodes~\cite{shockleyCurrentsConductorsInduced1938, sirkisCurrentsInducedMoving1966}, 
which can be picked up by connecting an RLC resonator to the electrodes, as demonstrated in standard Penning trap experiments~\cite{winelandPrinciplesStoredIon1975, bushevElectronsCryogenicPlanar2008,bohmanSympatheticCoolingTrapped2021, anCouplingTwoLasercooled2022, willSympatheticCoolingSchemes2022}.
However, in the context of QIP, the progress of single-electron image current detection in Paul traps is still hindered by motional frequency fluctuations that broaden the signal, making it harder to distinguish from the unavoidable thermal background.
This difficulty is deeply rooted in the fundamental design and operational constraints of the system.
First, achieving the strong readout coupling and steep magnetic field gradients required for QIP in Paul traps strongly motivates trap miniaturization~\cite{yuFeasibilityStudyQuantum2022,xu3DprintedMicroIon2025}, which typically increases the anharmonicity of the potential and, when combined with thermal motion, leads to frequency broadening.
Second, unlike the static magnetic confinement in Penning traps, Paul traps rely on an rf pseudopotential~\cite{leibfriedQuantumDynamicsSingle2003}. Consequently, amplitude fluctuations in the applied rf drive directly induce secular frequency drifts, and the inherent micromotion further complicates the electron dynamics.
To overcome these challenges, one approach is to lock the motional frequency using a parametric drive, which modulates the trapping potential at twice the original trapping frequency.
While this technique has been successfully demonstrated for electrons in Penning traps~\cite{winelandMonoelectronOscillator1973,tseng1bitMemoryUsing1999,tanSynchronizationParametricallyPumped1991,tanParametricallyPumpedElectron1993}, previous studies typically focused on steady-state dynamics~\cite{kimObservationHopfBifurcation2003,paparielloUltrasensitiveHystereticForce2016,leuchParametricSymmetryBreaking2016, eichlerParametricSymmetryBreaking2018,frimmerRapidFlippingParametric2019,xuSensitivityEnhancementNonlinear2024}, requiring detection wait times exceeding the damping time constant and thereby limiting the readout speed for QIP.
Furthermore, its applicability to the complex micromotion dynamics in Paul traps remains unexplored.

In this work, we present a comprehensive theoretical and numerical study of the parametric driving scheme for single-electron image current detection in Paul traps, where trap anharmonicity plays a beneficial role in stabilizing the electron motion.
Distinct from prior steady-state cases in Penning traps~\cite{winelandMonoelectronOscillator1973,tseng1bitMemoryUsing1999},
our work focuses on the transient regime before the system reaches the steady state and explicitly incorporates the inherent micromotion in Paul traps.
The shift towards the transient regime breaks the temporal constraints imposed by steady-state assumptions, enabling a fast readout scheme.
Specifically, we numerically demonstrate that, in the transient regime, an instantaneously applied drive results in a vanishing or small average image current at the corresponding frequency, whereas a slow, controlled ramp of the driving strength reliably generates robust signals, demonstrating strong resilience to realistic experimental noise.

\section{Detection scheme}
\begin{figure}
    \centering
    \includegraphics{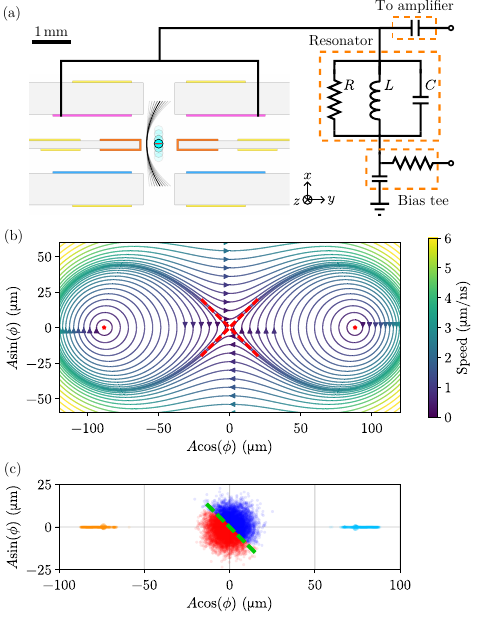}
    \caption{(a) Trap cross-section and detection circuit. 
    The three-layer PCB trap uses rf electrodes (orange) for radial ($xy$) confinement and dc electrodes (blue, pink and grounded yellow) for axial ($z$) confinement.
    Pink electrodes connect to a resonator with a bias-T to extract the single-electron image current.
    (b) Single-electron phase-space motion. 
    Streamlines map the electron trajectories, with local speed and direction encoded by color and arrows, respectively.
    The red stars denote stable solutions with damping. The red dashed lines mark the separatrix around center.
    (c) Phase-space scatter plot of the single-electron motion. 
    Initial phases (red/blue dots) evolve into final clusters (orange/light blue).
    The dashed green line marks the approximate theoretical separation threshold.
    }
    \label{fig1}
\end{figure}

Our goal is to achieve image current detection of single-electron motion by locking its motional frequency to half of the parametric drive frequency in the transient regime.
We choose the parametric driving over direct resonant driving to enable spectral filtering of the drive background and to facilitate future spin detection through phase bistability~\cite{tanSynchronizationParametricallyPumped1991,pengSpinReadoutTrapped2017}.
We consider a single electron trapped in a three-layer printed-circuit-board (PCB) trap, as shown in \figsub{fig1}{a}, with a distance of 0.635~mm between layers (half of that reported in~\cite{matthiesenTrappingElectronsRoomTemperature2021}). 
This reduced interlayer distance is fundamentally motivated by the induced image current mechanism, described by $I=ev/d_\mathrm{eff}$ for an electron of charge $e$ with velocity $v$ near the trap center~\cite{yuStrongCoherentIonelectron2024a,yuFeasibilityStudyQuantum2022}, given that $d_\mathrm{eff}$ roughly scales linearly with the trap size.
Two dc electrodes [pink in \figsub{fig1}{a}] on the top layer are connected to a high impedance resonator, represented by the RLC circuit inside the orange dashed box in \figsub{fig1}{a}.
This connection gives an effective distance of $d_\mathrm{eff}=4.8$~mm.
A bias-tee circuit connected afterwards is used for applying dc voltages to the electrodes during image current detection, and a capacitive connection to the amplifier represents the signal output.
When the electron is driven, the expected signal voltage at the output end is approximately proportional to $ A\omega R/d_\mathrm{eff}$, where $A$ and $\omega$ represent the amplitude and secular frequency of the electron motion, respectively, and $R$ is the on-resonance impedance of the resonator~\cite{brownGeoniumTheoryPhysics1986,yuFeasibilityStudyQuantum2022}.
In terms of trapping, the orange rf electrodes in the center PCB layer are supplied with ac voltages to generate radial confinement ($xy$-direction), while the blue and pink dc electrodes are applied with dc voltages to generate axial confinement ($z$-direction) together with the yellow grounded electrodes.

Under the pseudopotential approximation, and neglecting motional cross-coupling, the electron motion along the $x$-axis coupled to the resonator is modeled as
\begin{equation}\label{eq:x motion}
    \ddot{x}=-\omega_x^2x(1+\varepsilon\cos{\omega_{d}t})-\sum_{i=3}^{\infty}i{C_i}x^{i-1}-\gamma \dot{x},
\end{equation}
where $\omega_x$ is the secular frequency, $\gamma$ is the effective resonator-induced damping rate, $C_i$ denotes the $i$th-order anharmonic coefficients and $\varepsilon$ is the relative strength of the parametric drive at frequency $\omega_d$ applied via the rf electrodes.
Assuming an ansatz $x(t)=A(t)\cos(\omega_d t/2+\phi(t))$ near the principal parametric resonance ($\omega_d\approx2\omega_x$)~\cite{tseng1bitMemoryUsing1999,pengSpinReadoutTrapped2017}, we substitute it into \eq{eq:x motion} considering up to 6th order of the anharmonic terms and obtain
\begin{equation}\label{eq: A}
    \dot{A}=-\frac{\gamma}{2}A\left[1-\frac{\varepsilon\omega_x}{2\gamma}\sin{2\phi}\right],
\end{equation}
\begin{equation}\label{eq: phi}
    \dot{\phi}=\frac{1}{4}\varepsilon\omega_x\cos{2\phi}+\lambda_4A^2+\lambda_6A^4,
\end{equation}
where $\lambda_4=3C_4/2\omega_x$, $\lambda_6=15C_6/8\omega_x$.
The anharmonic terms introduce an amplitude-dependent frequency shift, limiting energy absorption from the parametric drive.

\begin{figure*}[tb]
    \centering
    \includegraphics{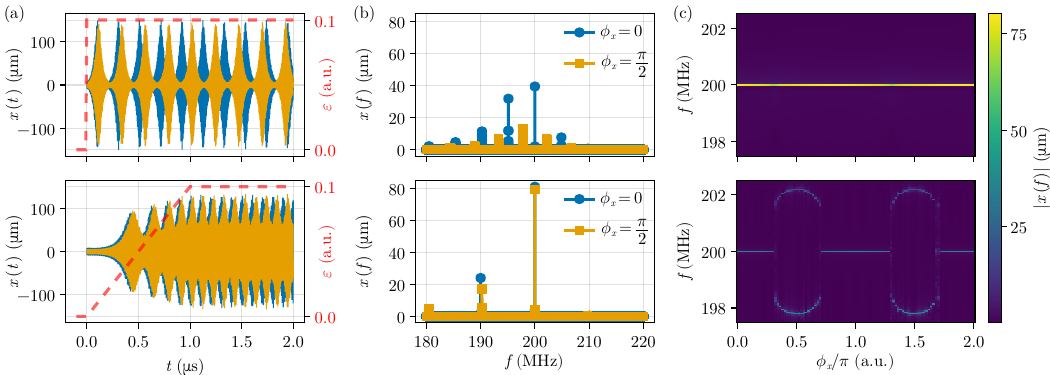}
    \caption{(a) Single electron motion along $x$-direction in the time domain comparing the instantaneous (top) with the ramp protocol (bottom). (b) Single electron motion along $x$-direction in the frequency domain. The red dashed lines in (a) show the relative driving strength $\varepsilon$. The blue and yellow lines in (a) and (b) correspond to the initial phase of the single electron in $x$-direction $\phi_x$ equal to $0$ and $\pi/2$, respectively. (c)~The frequency-domain amplitude of the single electron motion along $x$-direction for various $\phi_x$.  All top panels are with the relative driving strength $\varepsilon$ applied instantly at $t=0$~$\upmu$s, whereas all bottom panels with the relative driving strength $\varepsilon$ ramping up from $0$ to $0.1$ within $1$~$\upmu$s.}
    \label{fig2} 
\end{figure*}

As an example, \figsub{fig1}{b} shows the phase-space trajectories of the slow variables $A$ and $\phi$ with $\varepsilon=0.1$, $\omega_x=2\pi\times 200$~MHz, $\lambda_4= -4.08$~kHz/\unit{\um}$^{2}$, $\lambda_6= 6.78\times10^{-6}$~kHz/\unit{\um}$^{4}$ and the damping term ignored to reach the transient regime.
Here, $\lambda_4$ and $\lambda_6$ are obtained from the pseudopotential generated from the trap shown in \figsub{fig1}{a} with $\omega_{\rm{rf}}=2\pi\times1.452$~GHz.
The amplitude of the voltage applied to the rf electrodes for parametric driving with those parameters is two orders of magnitude smaller than the trap drive voltage (see Appendix \ref{supp: drive strength} for details).
In the transient regime, the trajectories, approximately forming closed orbits, are partitioned by a separatrix with a characteristic figure-eight topology, as shown in \figsub{fig1}{b}.
The separatrix asymptotically aligns with the dashed red lines in \figsub{fig1}{b}, where $\dot{\phi}=0$.
In the transient regime, the electron will oscillate around the attracting points (the red stars in \figsub{fig1}{b}), where $\dot{A}=\dot{\phi}=0$,  for time $\sim1/\gamma$ before relaxing to them~\cite{paparielloUltrasensitiveHystereticForce2016, strogatzNonlinearDynamicsChaos2024}.
When initial states are located within the left or right lobes of this figure-eight separatrix, the trajectories encircle only one of the star points, producing nonzero average amplitudes and deterministic average phases.
We characterize this motional state with a nonzero average amplitude and a deterministic average phase with respect to the corresponding frequency as being frequency-locked.
Conversely, for initial states located in regions outside the separatrix, the trajectories encompass the entire figure eight structure, enclosing both attractors and resulting in zero average amplitudes.
In this case, the induced image current will vanish at the expected frequency, preventing effective detection.
More severely, the phase of the motion is ill-defined with respect to the corresponding frequency, which is detrimental to the spin readout scheme that encodes the spin information into the phase of the electron motion~\cite{pengSpinReadoutTrapped2017}.

To guarantee a deterministic locking effect without waiting for the system to damp into a steady state, we propose to ramp up the relative drive strength $\varepsilon$.
This ramp is slow relative to the transient dynamics in the rotating frame, but remains much faster than the system damping time (e.g., $1$~\unit{\us} compared to the millisecond scale).
As $\varepsilon$ increases, a bifurcation occurs at the origin, causing the figure-eight separatrix to continuously expand outward.
Even for a very slow ramp of $\varepsilon$, the states will eventually cross the separatrix when the separatrix reaches them~\cite{caryAdiabaticinvariantChangeDue1986}.
For initial states located in the right (left) region of the separatrix, the trajectories will diabatically cross the separatrix and then adiabatically follow the right (left) attracting point if the separatrix expansion speed is slow enough.
For states located in the up (down) region, they will move to the right (left) side based on the trajectory shown in \figsub{fig1}{b} and then be captured by the right (left) side of the separatrix as those initially in the right (left) region.
To demonstrate this, we integrate \eq{eq: A} and \eq{eq: phi} to $t=20$~\unit{\us} with the same parameters as those used for \figsub{fig1}{b} except that $\varepsilon$ is linearly ramped up from $0$ to $0.1$ within $1$~\unit{\us}.
The central red and blue dots in \figsub{fig1}{c} represent the initial 10,000 random states for thermal motion at $4$~K.
Around the axis origin, they are roughly separated by the green dashed line.
After the ramp process, the red (blue) dots move to the left (right) side in the phase space, resulting in average values represented by orange (light blue) dots in \figsub{fig1}{c}.

\section{Three-dimensional dynamics}
To validate our model under realistic conditions, we numerically simulate the three-dimensional (3D) electron dynamics coupled to an RLC circuit using electric fields obtained from the actual three-layer trap geometry shown in \figsub{fig1}{a} via boundary element method (BEM)~\cite{singerColloquiumTrappedIons2010} (see Appendix \ref{supp: numerical} for numerical details). 
The dynamics can be modeled as
\begin{align}\label{eq: motion3D}
    m\ddot{\bm{r}}/q
    =&\bm{E}_\mathrm{dc}(\bm{r})+\bm{E}_\mathrm{rf}(\bm{r})\cos(\omega_\mathrm{rf}t+\phi_\mathrm{rf}) \nonumber\\
    &+\bm{E}_\mathrm{d}(\bm{r})\varepsilon(t)\cos(\omega_\mathrm{d}t+\phi_\mathrm{d})+L\dot{I}\bm{D}_\mathrm{inv}(\bm{r})\\
    LC\ddot{I}=&-\frac{L\dot{I}}{R}-I-q\bm{D}_\mathrm{inv}\cdot\dot{\bm{r}}
    \label{eq: motion3D current}
\end{align}
where $m$, $q$ and $\bm{r}$ denote electron mass, charge and position, respectively,
$R$, $L$ and $C$ are the effective parallel resistance, inductance and capacitance of the resonator, respectively,
$\bm{E}_\mathrm{dc, rf, d}$ denote the dc, rf and parametric drive fields, respectively,
$I$ is the current flowing through the effective inductance,
and $\bm{D}_\mathrm{inv}(\bm{r})=\bm{E}_w(\bm{r})/U_w$ characterizes the inverse of the effective electrode-electron distance with $\bm{E}_w(\bm{r})$ representing the electric field vector at position $\bm{r}$ for coupling wire potential $U_w$~\cite{yuStrongCoherentIonelectron2024a,winelandPrinciplesStoredIon1975}.
In the simulation, we assume the resonator quality factor to be $Q=1,000$ with a characteristic impedance of $Z_0=\sqrt{L/C}=300\, \Omega$, a resonant frequency $\omega_\mathrm{res}=1/\sqrt{LC}= 2\pi\times200$~MHz, resulting in an on-resonance impedance of $R=QZ_0=300$~k$\Omega$.
The values are conservative for a well-designed resonator at $4$~K~\cite{bohmanSympatheticCoolingTrapped2021, willSympatheticCoolingSchemes2022}.
The field amplitudes of $\bm{E}_\mathrm{dc, rf, d}$ are set by the rf drive frequency $\omega_\mathrm{rf} = 2\pi\times 1.452$~GHz based on the current experimental parameters of the trap drive resonance and secular frequencies $(\omega_x, \omega_y, \omega_z) = 2\pi \times (200, 173, 70)$~MHz.
The intentional detuning of $\omega_y$ from $200$~MHz prevents $y$-axis excitation by the $y$-component of the parametric drive $\bm{E}_\mathrm{d}$.

\begin{figure}[tb]
    \centering
    \includegraphics{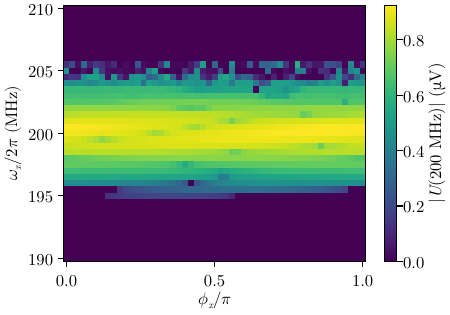}
    \caption{The induced voltage on the resonator at a frequency of $200$~MHz for various values of $\phi_x$ and $\omega_x$.}
    \label{fig3} 
\end{figure}

\figgsub{fig2}{a} shows the $x$-motion for initial phases $\phi_x=0$ (blue) and $\pi/2$ (yellow), comparing an instantaneous step of $\varepsilon=0.1$ (top) against a $1.0$-$\upmu$s ramp (bottom).  
Specifically, the system is initialized at temperature $T$ with $r_i=\sqrt{k_BT/m\omega_i^2}\cos\phi_i$ and $v_i=-\sqrt{k_BT/m}\sin\phi_i$ for $i=x,y,z$, where we set $\phi_y=\phi_z=\phi_\mathrm{rf}=0$. 
Under an instantaneous drive, frequency-locking to $\omega_d/2$ fails in the transient regime for initial phases $\phi_x \in (\pi/4, 3\pi/4) \cup (5\pi/4, 7\pi/4)$, the upper and lower regions delineated by red dashed lines in \figsub{fig1}{b}. In contrast, a ramping process suppresses these amplitude modulations [Fig.~\ref{fig2}(a), top] and locks the motion to $\omega_d/2$ with an enhanced amplitude across all values of $\phi_x$, as confirmed by the frequency spectra in \figsub{fig2}{b} and \fsub{fig2}{c}.
We then simulate the system dynamics for $\omega_x/2\pi \in [190, 210]$~MHz and $\phi_x \in [0, \pi]$ to evaluate the robustness against $x$-axis secular frequency fluctuations. 
Even when $\omega_x$ is detuned from $\omega_d/2=2\pi\times 200$~MHz by $2$\%, the circuit voltage induced by the electron motion in the $x$-direction maintains significant components ($\sim0.8$~\unit{\uV}) at $200$~MHz  as shown in \fig{fig3}, indicating strong robustness against experimental fluctuations of the rf amplitude~\cite{johnsonActiveStabilizationIon2016}.

\begin{figure}[tb]
    \centering
    \includegraphics{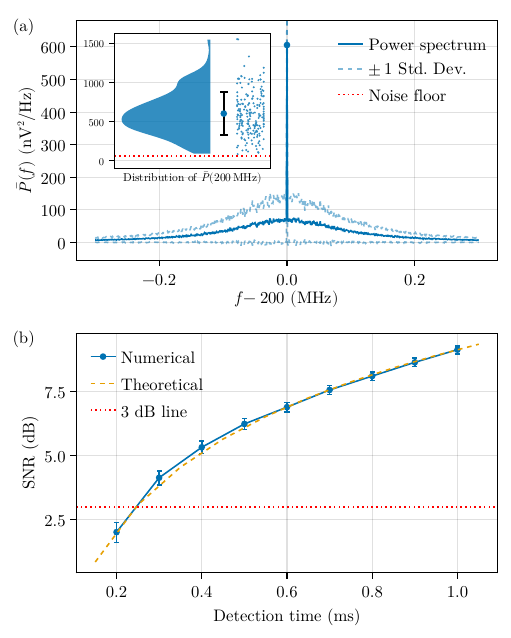}
    \caption{(a) Resonator voltage power spectrum. The solid and dashed lines represent the mean and standard deviation, respectively. Inset: power spetrum statistics at $200$~MHz, displaying the density distribution (shaded blue area), the mean and standard deviation (center dot with error bars), and the raw power spectrum values of individual trajectories (small dots). The dotted red line in the inset is the thermal noise floor.
     (b) SNRs of the voltage power signal of the image current with different detection times. 
    }
    \label{fig4} 
\end{figure}

\section{Quantifying noise impact}
To quantify the impact of noise, we employ stochastic differential equation simulations considering three noise sources: surface noise from trap electrodes, resonator Johnson noise, and amplitude fluctuations of both rf and parametric driving.
Based on trapped-ion experiments~\cite{brownnuttIontrapMeasurementsElectricfield2015}, we assume that the surface electric field noise scales as $S_E \propto \omega^{-1}d^{-2}T^{0.5}$, where $\omega$, $d$, and $T$ denote the frequency, electrode-to-particle distance, and temperature, respectively. Extrapolating a conservative baseline of $S_E = 10^{-12}\,\mathrm{V^2m^{-2}Hz^{-1}}$ (at $\omega/2\pi = 1\,\mathrm{MHz}$, $d=100$~\unit{\um}, $T=4$~K) to our system parameters ($d=431.8$~\unit{\um}, $T=4$~K) yields a noise level on the order of $10^{-16}\,\mathrm{V^2m^{-2}Hz^{-1}}$.
Since the system is only sensitive to a narrow frequency band, we approximate $S_E(\omega)$ as locally constant white noise. 
The field perturbation added to \eq{eq: motion3D} is $E_{n}=\sqrt{S_E(\omega)/2}\,\mathcal{N}(0,1/\delta t)$,
where $\mathcal{N}(0,1/\delta t)$ is a normal distribution centered around $0$ with varience $1/\delta t$, and $\delta t$ is the integration time step.
Similarly, Johnson noise is modeled by adding a random current $I_n=\sqrt{2k_BT/R}\,\mathcal{N}(0,1/\delta t)$ parallel to the RLC circuit in \eq{eq: motion3D current}, 
where $k_B$ is the Boltzmann constant, $T$ is the temperature~\cite{nyquistThermalAgitationElectric1928,gillespieMathematicsBrownianMotion1996,willSympatheticCoolingSchemes2022}.
Fluctuations in the rf electrode voltage, encompassing both the trapping and parametric driving fields, are modeled as a random walk process of the relative rf voltage, $R_U=U_\mathrm{rf}/U_\mathrm{rf}^\mathrm{ideal}$~\cite{willSympatheticCoolingSchemes2022}, where $U_\mathrm{rf}$ and $U_\mathrm{rf}^\mathrm{ideal}$ are the actual applied and ideal target voltages, respectively. We assume $R_U$ has an expectation value of $1$ and a standard deviation reaching $10^{-3}$ at $10$~ms.

To evaluate the signal-to-noise ratio (SNR) of the image current detection, we implement 200 random trajectories with the evolution time of $1$~\unit{ms}.
Using Fourier transfrom, we obtain the resonator voltage power spectrum as shown in \figsub{fig4}{a}, where the solid and dashed blue lines represent the mean and the standard deviation.
The inset of \figsub{fig4}{a} details the statistics of the resonator voltage power spectrum at the resonance frequency of $200$~MHz.
The SNR, plotted as a blue line in \figsub{fig4}{b}, is numerically extracted as
\begin{equation}
    \mathrm{SNR}=10\log\qty[\frac{\bar{P}(200\,\mathrm{MHz})-4k_BTR}{4k_BTR}],
\end{equation}
where $T=4$~K and $\bar{P}(200\,\mathrm{MHz})$ represents the average voltage power spectrum at $200$~MHz.
Results for various detection times are obtained by truncating the $1$-ms trajectories to the corresponding detection times.
The damping time constant of the system is $1/\gamma = m d_\mathrm{eff}^2/q^2R\approx 2.7$~ms~\cite{yuFeasibilityStudyQuantum2022}, where $d_\mathrm{eff} = 4.8$~mm.
The theoretical line in \figsub{fig4}{b} extrapolates the $1$-ms numerical SNR, assuming that the signal power $\bar{P}(200\,\mathrm{MHz})-4k_BTR$ scales linearly with the detection time.
As detailed in the inset of \figsub{fig4}{a}, despite a high average SNR and electron motion locked to $\omega_d/2$, the randomness of Johnson noise causes some individual trajectories to yield signal powers close to  the average noise level (the red dotted line in the inset). 
Because these inherent statistical fluctuations persist from the transient regime into the steady-state regime, this phenomenon offers a plausible explanation for the "mysterious bursts" observed in Ref.~\cite{bushevElectronsCryogenicPlanar2008} (see Appendix \ref{supp: signal bursts}), where they observed signal bursts that cannot be intuitively explained when they lowered the number of loaded electrons.

\section{Discussion}
In this work, we have investigated parametric driving in the transient regime for single-electron image current detection in a Paul trap. 
With detailed numerical simulation, we have discovered the importance of a slow controlled ramp of the driving field strength. 
The 3D simulation of the electron motion in the Paul trap with consideration of main noise sources demonstrates that our method is robust against realistic experimental noises under various initial conditions.
More importantly, our method fundamentally removes the temporal limitation applied by the steady-state assumptions and treats a typical side effect, the trap anharmonicity, as a beneficial resource to stabilize the electron motion.
Beyond single-electron detection, our framework can be extended to analyze the transient dynamics of the broader class of nonlinear oscillators under parametric driving~\cite{wangOIMOscillatorbasedIsing2019}, making it potentially applicable to systems such as nanomechanical resonators~\cite{braakmanForceSensingNanowire2019} and superconducting circuits~\cite{linJosephsonParametricPhaselocked2014,beaulieuCriticalityEnhancedQuantumSensing2025}.

\appendix

\section{Strength of the driving field}\label{supp: drive strength}
Here, we calculate the voltage ratio of the rf trapping drive to the parametric drive at the rf electrodes.
When the rf electrodes are applied with a voltage $V$ with frequency $\Omega$, the potential along $x$-direction under harmonic approximation should have the form of
\begin{equation}
    \frac{1}{2}mcVx^2\cos(\Omega t+\phi),
\end{equation}
where $m$ is the electron mass and $c$ represents a geometric constant that connects $V$ to the field strength.
The corresponding pseudopotential~\cite{leibfriedQuantumDynamicsSingle2003} is
\begin{equation}
    \frac{m(cV)^2}{4\Omega^2}x^2.
\end{equation}
For rf trapping drive, we have $\Omega=\omega_\mathrm{rf}$, $V=V_\mathrm{rf}$ and 
$$\frac{m(cV_\mathrm{rf})^2}{4\Omega^2}=\frac{m\omega_x^2}{2},$$
where $\omega_x$ is the $x$-direction secular frequency.
Here, we assume the $x$-direction confinement is provided by the rf trapping drive.
This then gives
\begin{equation}
    cV_\mathrm{rf}=\sqrt{2}\omega_x\omega_\mathrm{rf}.
\end{equation}
For parametric drive, we have $\Omega=\omega_d$, $V=V_\mathrm{d}$ and 
\begin{equation}
    cV_d=\varepsilon\omega_x^2.
\end{equation}
Therefore, the voltage ratio of the rf drive to the parametric drive is 
\begin{equation}
    \frac{V_\mathrm{rf}}{V_d}=\frac{\sqrt{2}\omega_\mathrm{rf}}{\omega_x\varepsilon}.
\end{equation}
For $\varepsilon=0.1$, $\omega_\mathrm{rf}=(2\pi)\,1.452$~GHz, $\omega_x=(2\pi)\,200$~MHz, we have $V_\mathrm{rf}/V_d=102.67$, which corresponds to $40.2$~dB in power assuming the same impedance.
This large ratio demonstrates that the required voltage for the parametric drive ($V_d$) is relatively small compared to the rf drive for trapping ($V_\mathrm{rf}$). 
Therefore, the $400$~MHz parametric drive may not require resonant voltage enhancement from a resonator, allowing the drive to be superimposed onto the primary $1.452$~GHz trapping field at the rf electrodes.

\section{Details on numerical simulation}\label{supp: numerical}
The electric field data for numerical trajectory simulation are obtained by simulating the fields of each electrode with BEM in the region of interest (ROI) and then applying a linear combination of all the fields based on the desired frequencies of the electron motion.
The geometry of the electrodes is adopted from Ref.~\cite{matthiesenTrappingElectronsRoomTemperature2021}, while the distance between PCB boards is shrinked by a factor of $2$.
The ROI is selected to be from -$150$ to $150$~\unit{\um} along the $x$ and $y$ directions, and $-25$ to $25$~\unit{\um} along the z-direction, with $1$~\unit{\um} resolution, for a reasonable computational time. 
The field is then interpolated by fitting the spherical harmonic basis up to the $6$th order. The effective electron-electrode distance $d_\mathrm{eff}$ can also be extracted using the electric field generated by the pickup electrodes, which is $4.8$~mm at the trap center in the $x$-direction.

The numerical simulations of the system dynamics are carried out using adaptive step-size solvers to ensure both computational efficiency and accuracy. 
We verified the convergence of our results by varying the step-size control options, which evaluate the local absolute and relative errors ($abstol$ and $reltol$) at each integration step. 
The 1D simulations presented in \figsub{fig1}{c} are implemented in Python using \verb|Tsit5| (a 5th-order explicit Runge-Kutta method)~\cite{tsitouras2011runge} with $abstol=reltol=10^{-10}$. 
For 3D simulations, we utilize the Julia package \verb|DifferentialEquations.jl|~\cite{rackauckas2017differentialequations}. 
Specifically, ordinary differential equations (ODEs) are integrated using \verb|Vern9| (a modified 9th-order explicit Runge-Kutta method)~\cite{verner2010numerically}, while stochastic differential equations (SDEs) are solved with \verb|SkenCarp| (a solver sutible for additive noise problems )~\cite{kennedy2001additive,rackauckas2017differentialequations}. 
The data for Figs.~\ref{fig2} and \ref{fig3} are obtained using \verb|Vern9| with $abstol=reltol=10^{-10}$, while the SDE simulations for \fig{fig4} are performed with $abstol=reltol=10^{-9}$.
The phase-space trajectory plot in \figsub{fig1}{b} is generated using the \verb|streamplot| function in the Python package \verb|matplotlib|~\cite{Hunter:2007}. 
The stremlines are computed on a spatial grid ranging from $-120$ to $120$~\unit{\um} with $1001$ points in each dimension.
To enhance visual clarity, the final displayed plot is cropped to a window of $-60$ to $60$~\unit{\um} along the $A\sin(\phi)$ axis.

\begin{figure}[tb]
    \centering
    \includegraphics{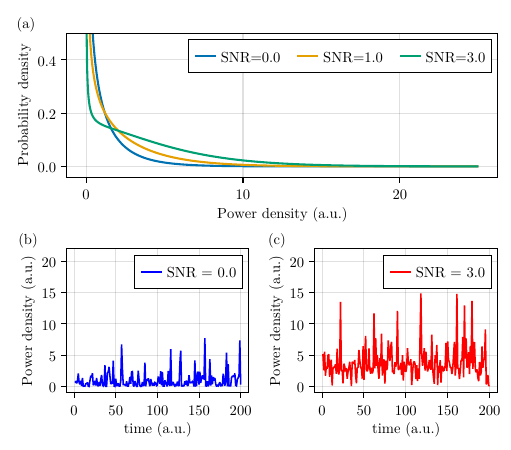}
    \caption{(a) Probability density distribution for the power density at resonant frequency with different SNR. (b) and (c) Power density at resonant frequency over time for $\mathrm{SNR}=0.0$ and $\mathrm{SNR}=3.0$, respectively.}
    \label{suppfig: SNR bursts}
\end{figure}

\section{Bursts of voltage power }\label{supp: signal bursts}
Here we explain the voltage power bursts at resonance frequency by looking at the random variable in theory.
The voltage power density (spectrum) can be expressed as
\begin{equation}
    P(f)=\qty[V_s(f)+V_n(f)]^2,
\end{equation}
where $f$ is the frequency, $V_s$ is the signal voltage amplitude density induced by the electron motion and $V_n$ is the voltage amplitude density of the Johnson noise which satisfies normal distribution $\mathcal{N}(0,4k_BTR)$~\cite{nyquistThermalAgitationElectric1928} when $f$ is equal to the resonant frequency $f_r$ of the resonator.
The voltage power density satisfies the non-central chi-squared distribution~\cite{papoulisRandomVariablesStochastic1991}, as shown in \figsub{suppfig: SNR bursts}{a}. 
Even when $\mathrm{SNR}=3.0$, the power density still has a large probability with low power density values similar to the $\mathrm{SNR}=0.0$ case.
As shown in \figsub{suppfig: SNR bursts}{b} and \fsub{suppfig: SNR bursts}{c}, the time series of the power density for the different cases still share the same bursting phenomena with different power density amplitudes. 
At each time data point in \figsub{suppfig: SNR bursts}{b} and \fsub{suppfig: SNR bursts}{c}, the value represents an independent sample of this power density that satisfies the non-central chi-squared distribution shown in \figsub{suppfig: SNR bursts}{a}.

\bibliography{references}

\end{document}